\documentclass[twocolumn, tighten,preprint]{aastex63}
\usepackage{apjfonts}
\usepackage{color}
\usepackage{amsmath, mathrsfs}
\usepackage{amssymb}

\DeclareMathAlphabet{\mathbi}{OT1}{ptm}{bx}{it}
\SetMathAlphabet\mathbi{bold}{OT1}{ptm}{bx}{it}

\def\cblue{}

\begin{document}

\title{\bf\large A Pixon-Based Method for Reverberation-mapping Analysis in Active Galactic Nuclei}

\author[0000-0001-5841-9179]{Yan-Rong Li}
\author{Ming Xiao}
\affiliation{Key Laboratory for Particle Astrophysics, Institute of High 
Energy Physics, Chinese Academy of Sciences, 19B Yuquan Road, 
Beijing 100049, China; \href{mailto:liyanrong@mail.ihep.ac.cn}{liyanrong@mail.ihep.ac.cn}}
\author[0000-0001-9449-9268]{Jian-Min Wang}
\affiliation{Key Laboratory for Particle Astrophysics, Institute of High 
Energy Physics, Chinese Academy of Sciences, 19B Yuquan Road, 
Beijing 100049, China; \href{mailto:liyanrong@mail.ihep.ac.cn}{liyanrong@mail.ihep.ac.cn}}
\affiliation{School of Astronomy and Space Science, University of Chinese Academy of Sciences, 
19A Yuquan Road, Beijing 100049, China}
\affiliation{National Astronomical Observatories of China, Chinese 
Academy of Sciences, A20 Datun Road, Beijing 100012, China}

\begin{abstract}
We present an alternative method for reconstructing a velocity-delay map in reverberation mapping (RM) based on the pixon algorithm
initially proposed for image reconstruction by \citeauthor{Pina1993}. The pixon algorithm allows for a variable pixon 
basis to adjust resolution of each image pixel according the information content in that pixel, which therefore enables the algorithm 
to make the best possible use of measured data.
The final optimal pixon basis functions would be those {\cblue that} minimize the number of pixons while still {\cblue providing} acceptable descriptions to data within the accuracy {\cblue allowed} by noises. We adapt the pixon algorithm to RM 
analysis and develop a generic framework to implement the algorithm. Simulation tests and comparisons with the widely used 
maximum entropy method demonstrate the feasibility and high performance of our pixon-based RM analysis.
This paper serves as an introduction to the framework and the application to 
velocity-unresolved RM. An extension to velocity-resolved cases will be presented in a companion paper.
\end{abstract}

\section{Introduction}
The reverberation mapping (RM) technique  in active galactic nuclei (AGNs) measures 
responses of variations of reprocessing emissions to those of driving emissions (\citealt{Blandford1982, Peterson1993}).
The responses as functions of time delay and wavelength (or velocity) generally decode information about 
 geometry and kinematics of the reprocessing regions.
Over the past three decades, the great successful application of RM to broad emission lines in AGNs
had significantly advanced our understanding {\cblue{of}} the gaseous regions (the so-called broad-line regions, BLRs) 
responsible for broad emission lines (e.g., \citealt{Kaspi2000, Bentz2013, Du2014, Du2016}) and, most importantly, yielded an 
efficient mass estimate of the central supermassive black holes (SMBHs, e.g., \citealt{Peterson2004, Peterson2014}). 
Only with this mass estimate were systematic studies on cosmological evolution of SMBHs made possible
(e.g., \citealt{Marconi2004, Wang2009, Li2011, Li2012, Kelly2013, Shankar2013}). 

Another important application of RM is diagnosing structures of the accretion disk (e.g., 
\citealt{Sergeev2005,Edelson2019, Cackett2020}) 
and hot corona, a region believed to be located around the inner accretion disk and to mainly emit X-rays 
(e.g., \citealt{Reynolds1999, Alston2020}). The key finding toward this line, mainly revealed from the {\it Swift} intense 
accretion disk RM survey,
was that UV/optical interband lags are consistent with the lag-wavelength relationship ($\tau\propto \lambda^{4/3}$)
predicted by the standard accretion disk model. However, the amplitudes of interband lags are generally larger by a factor 
of $\sim2$ than the anticipated values (e.g., \citealt{Edelson2019} and references therein). This had spurred extensive 
{\cblue investigations} over the accretion disk model itself and other physical processes possibly involved 
(e.g., \citealt{Dexter2011,Hall2018, Korista2019, Sun2020}).

In the coming era of massive time-domain surveys (e.g., the Zwicky
Transient Facility, \citealt{Bellm2019}; the Large Synoptic Survey Telescope, \citealt{Ivezic2008}), the RM technique 
is expected to {\cblue provide} much more profound insight into geometry and kinematics 
of gaseous environments surrounding the vicinity of the central SMBHs.
This calls for developing alternative sophisticated methods {\cblue rather than using} 
the traditional {\cblue simple} cross-correlation method
to make the best use of information contained in measured data.
Indeed, previously there had been several methods proposed/developed for this purpose, including 
the maximum entropy technique (\citealt{Horne1994}), the regularized linear inverse method (\citealt{Krolik1995}),
the SOLA method (\citealt{Pijpers1994}), and other forward methods with a presumption of {\cblue AGN variability} models 
(\citealt{Zu2011, Li2016, Starkey2016}). Among these methods, the maximum entropy technique
had found several applications (e.g, \citealt{Horne2003, Horne2020, Bentz2010, Grier2013, Xiao2018, Xiao2018b}) owing 
to its simple underlying principle and flexibility 
in describing the data and suppressing the noise propagation.

In the maximum entropy method (MEM), the entropy acts as a prior probability to a specific solution from a Bayesian 
viewpoint. In the absence of any additional information, this entropy prior states that the flattest solutions are the most 
probable. However, in reality, we expect that true meaningful solutions deviate from flatness and 
show structures. In this sense, the default flatness is no longer an ideal prior, which instead should be assigned according to the
information content in measured data and truly maximized with the best solutions%
{\footnote{It is worth mentioning that in the MEM one can design some sort of default prior images appropriately changing 
with the local image solutions, which partially surmounts the issue of the flat prior (\citealt{Skilling1989,Horne1994}). }. 
On the other hand, there are usually some 
parts in the solutions that are blank and have no information content. These blank parts should be assigned
low weights or excluded from degrees of freedom. Considerations such as these led \cite{Pina1993}
to introduce the pixon concept for image reconstruction. 
A variety of astronomical applications demonstrated high performance of the pixon method 
and its great capability in noise suppressing and robust rejection of spurious sources in image 
reconstructions (\citealt{Metcalf1996, Putter1996, Dixon1997, Putter1999, Eke2001}).
Briefly speaking, a pixon is a generalized pixel and basic unit for image reconstruction.
Pixons are able to vary in size to adapt to the measured data so as to make the information content 
flat across the inferred solutions in terms of pixons. Smallest pixons are used to describe regions with the highest
information content, whereas large pixons are used for low signal-to-noise ratio {\cblue (S/N)}  regions.
The ultimate goal of the pixon method is to find the fewest pixons that still yield 
acceptable fits to the measured data. This is generally implemented in an iterative way,
and the termination point is well defined.

The essence of RM analysis is deconvolving the integral equation
\begin{equation}
F_l(v, t) = \int \Psi(v, \tau) F_c(t-\tau) d\tau,
\label{eqn_rm}
\end{equation}
where $F_c(t)$ and $F_l(v, t)$ are observables that represent variations of the continuum and emission line at 
time $t$ and velocity $v$, respectively, $\Psi(v, \tau)$ is the transfer function, and $\tau$ is the time lag.
This is in analogy with an image reconstruction process, meaning that the pixon method 
can also be adapted to RM analysis. An unimportant difference is that in image reconstructions the transfer functions 
are known and the true images are to be determined; by contrast, in RM the situation is just the reverse, namely, 
the true images (or the continuum light curves) are known and the transfer functions are to be determined.
However, this does not affect our application of the pixon concept to RM analysis at all.

In this paper, we adapt the pixon method and develop a mathematical framework
to implement the pixon concept in RM analysis. We are only concentrated on the light curves of 
velocity-integrated emission-line fluxes (namely, velocity unresolved) so that the integral equation is simplified to 
\begin{equation}
F_l(t) = \int \Psi(\tau) F_c(t-\tau) d\tau.
\label{eqn_rm1d}
\end{equation}
It is straightforward to extend the current framework to velocity-resolved RM analysis.

The paper is organized as follows.
Section 2 describes the methodology of the pixon concept and application to 
RM analysis. Section 3 performs a number of simulation tests to verify the validity and 
feasibility of our pixon-based RM analysis. In Section 4, we make a comparison to the MEM.
Discussion and conclusion are given in Section 5.

\section{Methodology}
In this section, we first briefly describe the pixon concept and the necessary formulae 
following the denotations in \cite{Metcalf1996}. We then adapt the pixon method 
to RM analysis. Considering that observed light curves generally are irregularly 
sampled, we present two approaches for reconstructing continuum light curves on
evenly spaced times. The first approach is based on the pixon concept, and 
the other is based on the damped random walk (DRW) process.

\subsection{The Pixon Concept}
According to Bayes' theorem, the posterior probability
of the reconstructed image ($I$) and model ($M$) given the observed data ($D$)
is written as
\begin{equation}
P(I, M|D) = \frac{P(D|I, M)P(I|M)P(M)}{P(D)}\propto P(D|I, M) P(I|M),
\end{equation}
where $P(D|I, M)$ is likelihood probability, $P(D)$ is the marginal likelihood or evidence, 
$P(I|M)$ is the image prior that does not depend on the data, and $P(M)$ is the model prior that is 
generally presumed to be uniform. 

By assuming that the measurement {\cblue errors} are independent and Gaussian, the likelihood probability reads 
\begin{equation}
P(D|I, M) \propto \exp \left(-\frac{\chi^2}{2}\right),
\label{eqn_likelihood}
\end{equation}
where $\chi^2$ is the sum of the squares of the standardized residuals (see below). 
The image prior can be quantified by considering the simple counting argument, namely, distributing 
$N$ indistinguishable photons randomly among $n_{\rm pixon}$ pixons (or cells). 
By denoting the number of photons in pixon $i$ to be $N_i$, the image prior 
of a particular distribution is 
\begin{equation}
P(I|M) = \frac{N!}{n^N\prod_i N_i!} \propto \frac{1}{(n_{\rm pixon})^N}\exp\left(S\right),
\label{eqn_image_prior}
\end{equation}
where Stirling's approximation to factorials is used to derive the far right-hand side and
\begin{equation}
S = -\sum_{i=1}^{n_{\rm pixon}}\frac{N_i}{N}\ln\frac{N_i}{N}.
\end{equation}
Here we note that the summation runs over pixons instead of pixels.

As can be seen, the image prior probability increases as the number of pixons decreases. 
This is why the pixon method seeks to find the fewest number of pixons.
The exponential term in Equation~(\ref{eqn_image_prior}) can be deemed to be a sort of
entropy as in the MEM. With Equations~(\ref{eqn_likelihood}) and (\ref{eqn_image_prior}), 
the posterior probability is given by
\begin{equation}
P(I, M|D)\propto \frac{1}{(n_{\rm pixon})^N}\exp\left[-\frac{1}{2}\left(\chi^2-2S\right)\right].
\label{eqn_post}
\end{equation}
In practice, the ratio $N_i/N$ can be recast into $I_i/I_{\rm tot}$, where $I_i$ is the physical quantity (to be solved) in 
the $i$-th pixon and $I_{\rm tot}=\sum_i I_i$. However, there is not a clear way to calculate the total photon number
$N$ considering that we are concentrated on solving the transfer functions. It is thus not
straightforward to directly explore this posterior probability defined above. 
We note that the factor $1/(n_{\rm pixon})^N$ is maximized provided that the pixon number is as {\cblue small} as possible.
As a result, optimizing the above posterior probability is indeed bound to maximize the exponential term in
Equation~(\ref{eqn_post}), as well as find the fewest number of pixons.

To implement the pixon concept, \cite{Pina1993} introduced the fractal pixon basis,
which is a family of chosen functions with various widths.
The reconstructed image $I(x)$ is then represented as a convolution of a pseudo-image $I^{(p)}(x)$ with 
a pixon basis function appropriate at pixel $x$,
\begin{equation}
I(x) = \int I^{(p)}(y) K_x\left(\frac{y-x}{\delta_x}\right) dV_y,
\end{equation}
where $\delta_x$ is the width of the pixon basis function $K_x$, which is normalized such that 
\begin{equation}
\int K_x\left(\frac{y}{\delta_x}\right) dV_y  = 1.
\end{equation}
In this implementation, each pixon shares some fraction of signals from its adjacent pixons
and there are {\cblue no} hard edges between pixons. The pixon density at $i$-pixel is 
\begin{equation}
f_i = \left[\int  k_i\left(\frac{y}{\delta_i}\right)dV_y \right]^{-1},
\label{eqn_f}
\end{equation}
where $k_i$ is the pixon basis function normalized to 1 at $y=0$.
The number of pixons is the sum of $f_i$ over all pixels, namely (\citealt{Metcalf1996}),
\begin{equation}
n_{\rm pixon} = \sum_{i=1}^{n_{\rm pixel}} f_i,
\end{equation}
where $n_{\rm pixel}$ is the number of pixels.
The entropy term in Equation~(\ref{eqn_post}) can be recast into the summation over pixels as
(\citealt{Metcalf1996})
\begin{equation}
S = -\sum_{i=1}^{n_{\rm pixel}}\frac{N_i}{N}\ln\frac{N_i}{f_iN}
 \approx  -\alpha \sum_{i=1}^{n_{\rm pixel}}\frac{N_i}{N}\ln\frac{N_i}{N},
\end{equation}
where on the right-hand side $N_i$ refers to $i$-pixel and again $N_i/N$ can be replaced 
with $I_i/I_{\rm tot}$ to make it calculable, and 
\begin{equation}
\alpha = \frac{\ln n_{\rm pixon}}{\ln n_{\rm pixel}}.
\end{equation}
The entropy term is indeed not important in the pixon method because 
pixons already induce effective smoothing to make solutions as flat as possible (\citealt{Metcalf1996}).

\begin{figure}[t!]
\centering 
\includegraphics[width=0.45\textwidth]{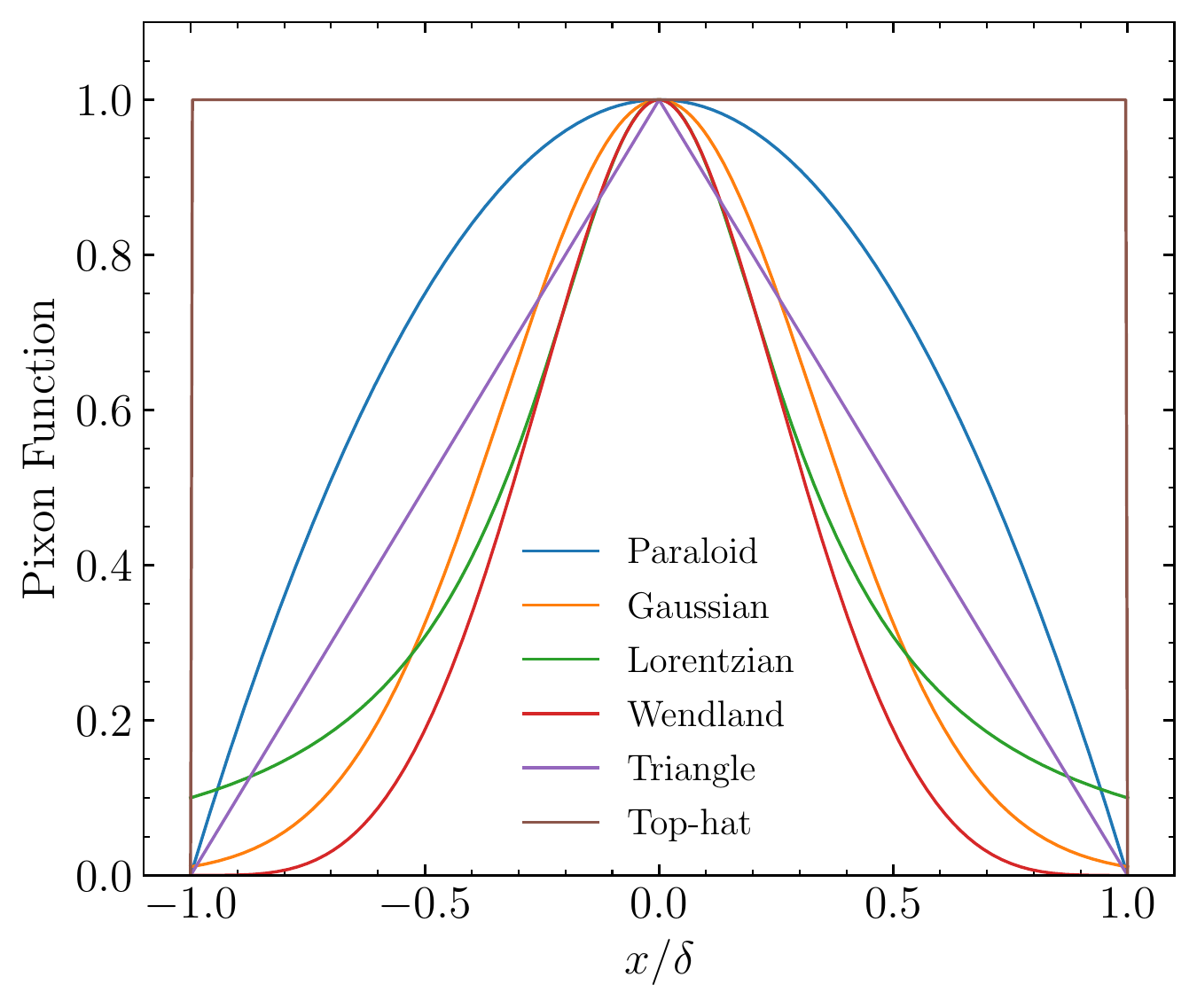}
\caption{Shapes of pixon basis functions listed in Table~\ref{tab_pixon}. The pixon functions are truncated at $x=\pm\delta$.
\label{fig_pixon}}
\end{figure}
\begin{deluxetable}{lll}
\renewcommand\arraystretch{1.5}
\tablecolumns{3}
\tabletypesize{\footnotesize}
\tablecaption{Pixon shapes and normalizations.\label{tab_pixon}}
\tablehead{
\colhead{~~~~~~~Name~~~~~~~} &
\colhead{~~~~~~~Shape~~~~~~~} &
\colhead{~~~~~~~Normalization~~~~~~~}
}
\startdata
Paraboloid & $\displaystyle 1-\frac{x^2}{\delta^2}$  & $\displaystyle\frac{3}{4\delta}$  \\
Gaussian  & $\displaystyle\exp\left(-\frac{1}{2}\frac{9x^2}{\delta^2}\right)$  & $\displaystyle\left[\sqrt{2\pi}{\rm erf}\left(\frac{3}{\sqrt{2}}\right)\frac{\delta}{3}\right]^{-1}$ \\
Lorentzian & $\displaystyle\frac{1}{1+9x^2/\delta^2}$  & $\displaystyle\left[\frac{2\delta}{3}\tan^{-1}(3)\right]^{-1}$\\
Wendland & $\displaystyle \left(1-\frac{|x|}{\delta}\right)^4\left(4\frac{|x|}{\delta}+1\right)$  & $\displaystyle\frac{3}{2\delta}$ \\
Triangle & $\displaystyle 1-\frac{|x|}{\delta}$     & $\displaystyle\frac{1}{\delta}$ \\
Top hat  & 1                                        & $\displaystyle\frac{1}{2\delta}$
\enddata
\tablecomments{All pixon functions are truncated at $|x|=\delta$, beyond which the function values vanish.}
\end{deluxetable}

There are no restrictions on selection of the pixon basis functions {\it a priori}. 
Truncated paraboloids were widely adopted in applications of the pixon concept (\citealt{Pina1993, Metcalf1996, Dixon1997})
largely because of the computational demands. The other functions, such as truncated Gaussians, 
were also adopted (\citealt{Eke2001}). We list 
several pixon basis functions and their normalizations in Table~\ref{tab_pixon} and plot the corresponding shapes 
in Figure~\ref{fig_pixon}. In most cases, the results are not 
sensitive to the selection of pixon basis functions. However, these pixon basis functions with sharp edges (e.g., top hats) would be more likely 
to produce unsmoothed features in reconstructed images, which are not always desirable.

\begin{figure*}[th!]
\centering 
\includegraphics[width=0.9\textwidth]{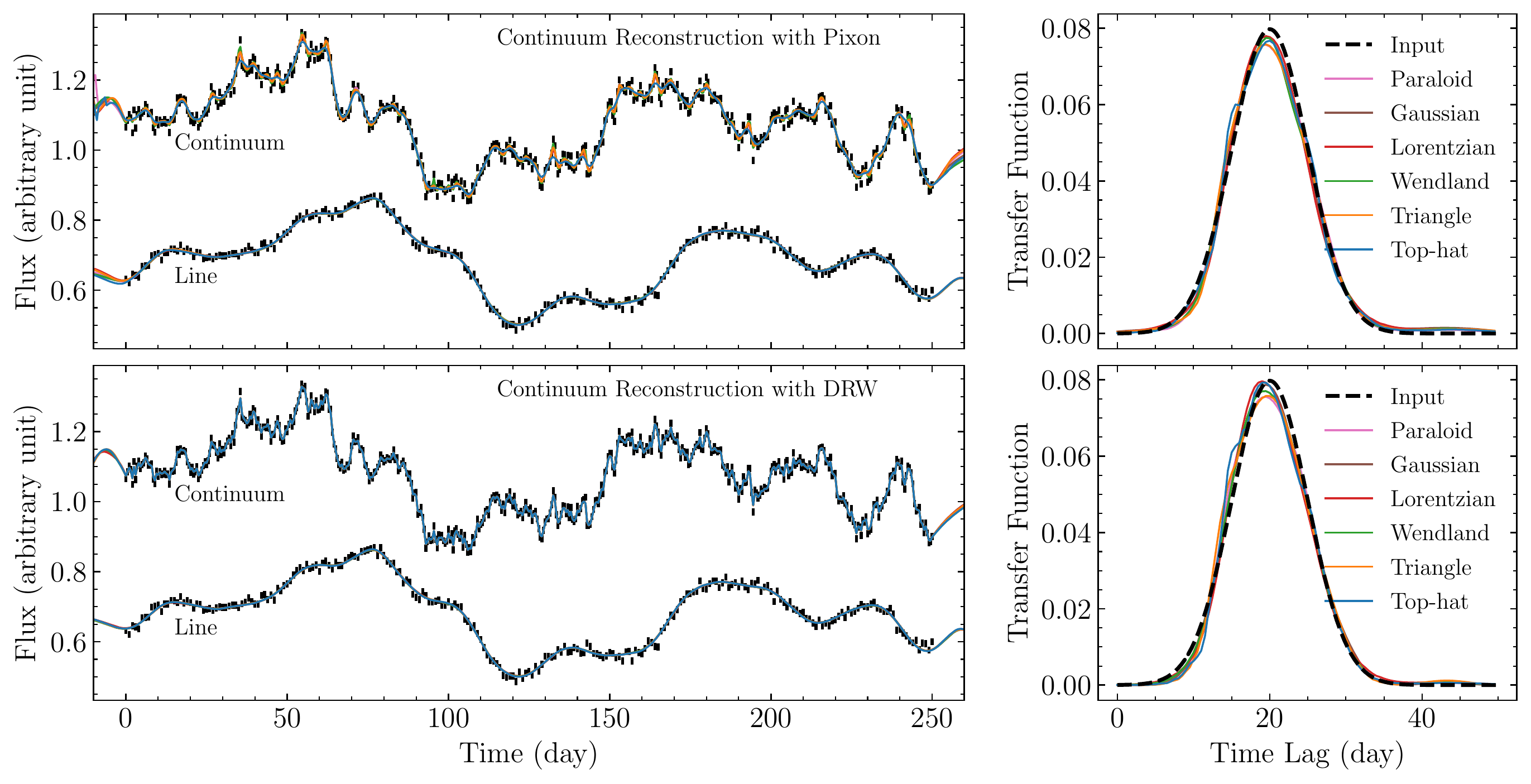}
\caption{Validity tests of our pixon-based approach for RM analysis. Top panels show results for the case of 
continuum reconstruction using the pixon concept, and 
bottom panels are for the case of continuum reconstruction using DRW. In each case, the left panel shows the simulated 
light curves of continuum and emission line
(black error bars) and their reconstruction (solid lines), and the right panel shows reconstructed transfer functions 
with different pixon basis functions listed in Table~\ref{tab_pixon}.
Black solid lines represent the input transfer function. 
\label{fig_validity_tests}}
\end{figure*}

\subsection{Application to Reverberation Mapping}
\label{sec_cont}
In RM analysis, the quantity to be determined is the transfer function $\Psi(\tau)$ in Equation~(\ref{eqn_rm1d}).
Following the nomenclature introduced in the preceding section, we use the term ``image'' to represent 
the transfer function and $I_i$ to denote its value at $i$-pixel.
The {\cblue chi square} for the emission-line data $D_l$ is 
\begin{equation}
\chi^2_{l} = \sum_j \frac{(F_{l,j} - D_{l, j})^2}{\sigma_{l, j}^2},
\end{equation}
where $D_{l,j}$ and $\sigma_j$ are the measured flux and error of the emission line at time $t_j$, respectively, 
and $F_{l,j}$ is the reconstructed flux using Equation~(\ref{eqn_rm1d}).

To implement the convolution of Equation~(\ref{eqn_rm1d}), one needs to reconstruct 
continuum fluxes from the usually irregularly sampled continuum light curve.
We present two approaches for this purpose: one is based on the pixon method itself, and the other 
is based on the DRW process.

\subsubsection{Continuum Reconstruction Using the Pixon Concept}
\label{sec_cont_pixon}
Similar to the above-mentioned pixon concept, we can also use pixons to 
reconstruct continuum light curve as 
\begin{equation}
F_c(t) = \int F_{c}^{(p)}(t') K_t\left(\frac{t - t'}{\delta_t}\right) dt',
\end{equation}
where $F_{c}^{(p)}$ is the pseudo-continuum to be determined, $K_t$ is the pixon function appropriate at time $t$
and $\delta_t$ is the width of the pixon function.
The chi square of continuum reconstruction is 
\begin{equation}
\chi^2_c = \sum_j \frac{(F_{c,j} - D_{c,j})^2}{\sigma_{c, j}^2},
\end{equation}
where $D_{c,j}$ and $\sigma_{c, j}$ are the measured flux and error of the continuum at time $t_j$.
For simplicity, we use a uniform pixon size for continuum reconstruction.

\begin{figure*}[th!]
\centering 
\includegraphics[width=0.9\textwidth]{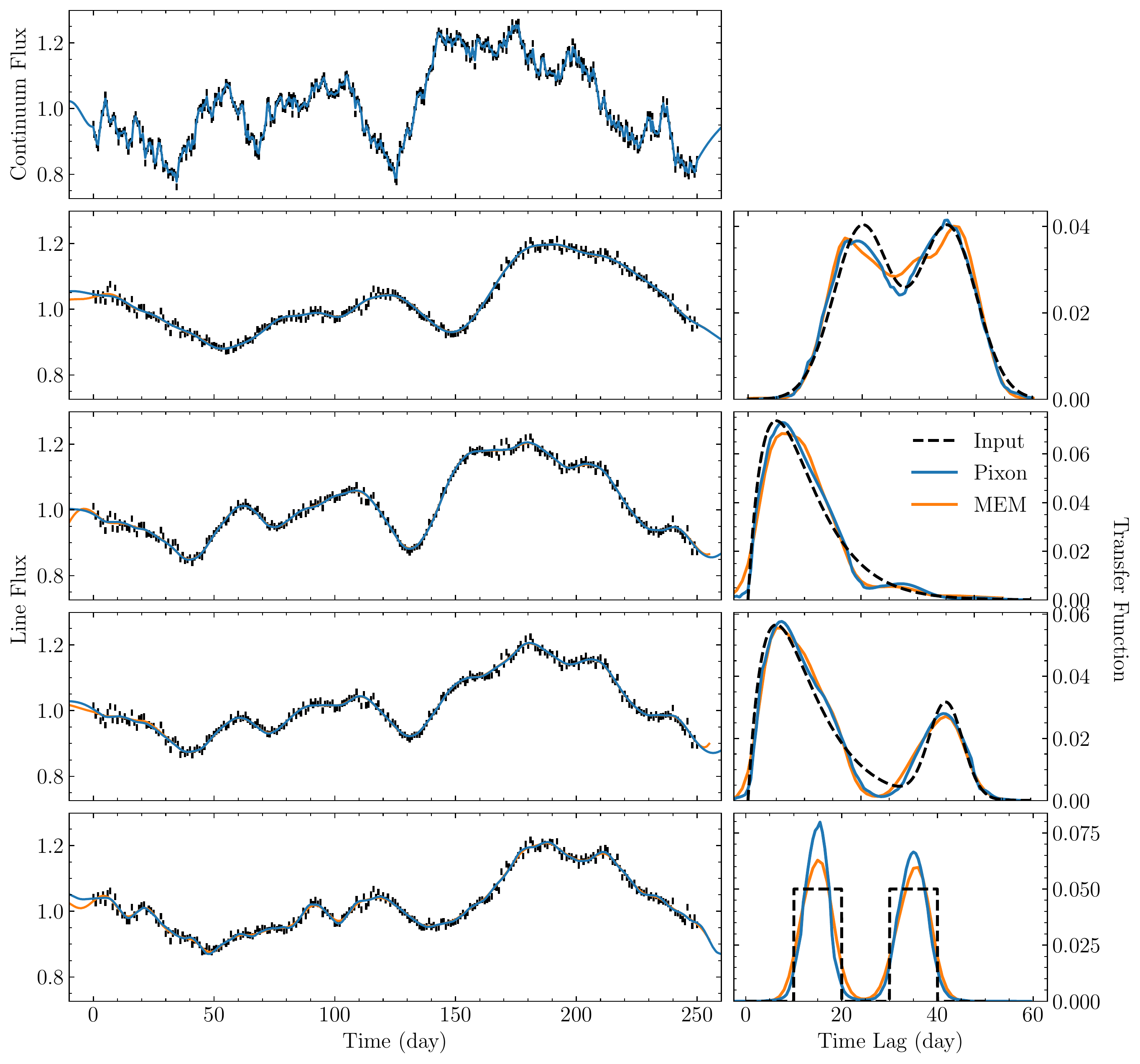}
\caption{Validity tests of our pixon-based approach with various shapes of input transfer functions: two highly blending 
Gaussians, a fast-rising peak with a long-descending tail,  a fast-rising peak with a long-descending tail plus a Gaussian, and two displaced top hats (top to bottom). The top left panel shows the generated continuum light curve. Four panels underneath show simulated light curves 
of the emission line using the transfer functions plotted in the right panels (black dashed lines). 
Blue solid lines represent reconstructions to the light curves and recovered transfer functions. 
The pixon basis functions are set to Gaussian. 
In the top left panel, only the reconstruction to the continuum light curve for the case of two highly blending Gaussians is plotted.
For the sake of comparison, the reconstructions to light curves and transfer functions by the MEM are also superposed (yellow solid lines).
\label{fig_tranfuns_tests}}
\end{figure*}

\begin{figure}[th!]
\centering 
\includegraphics[width=0.33\textwidth]{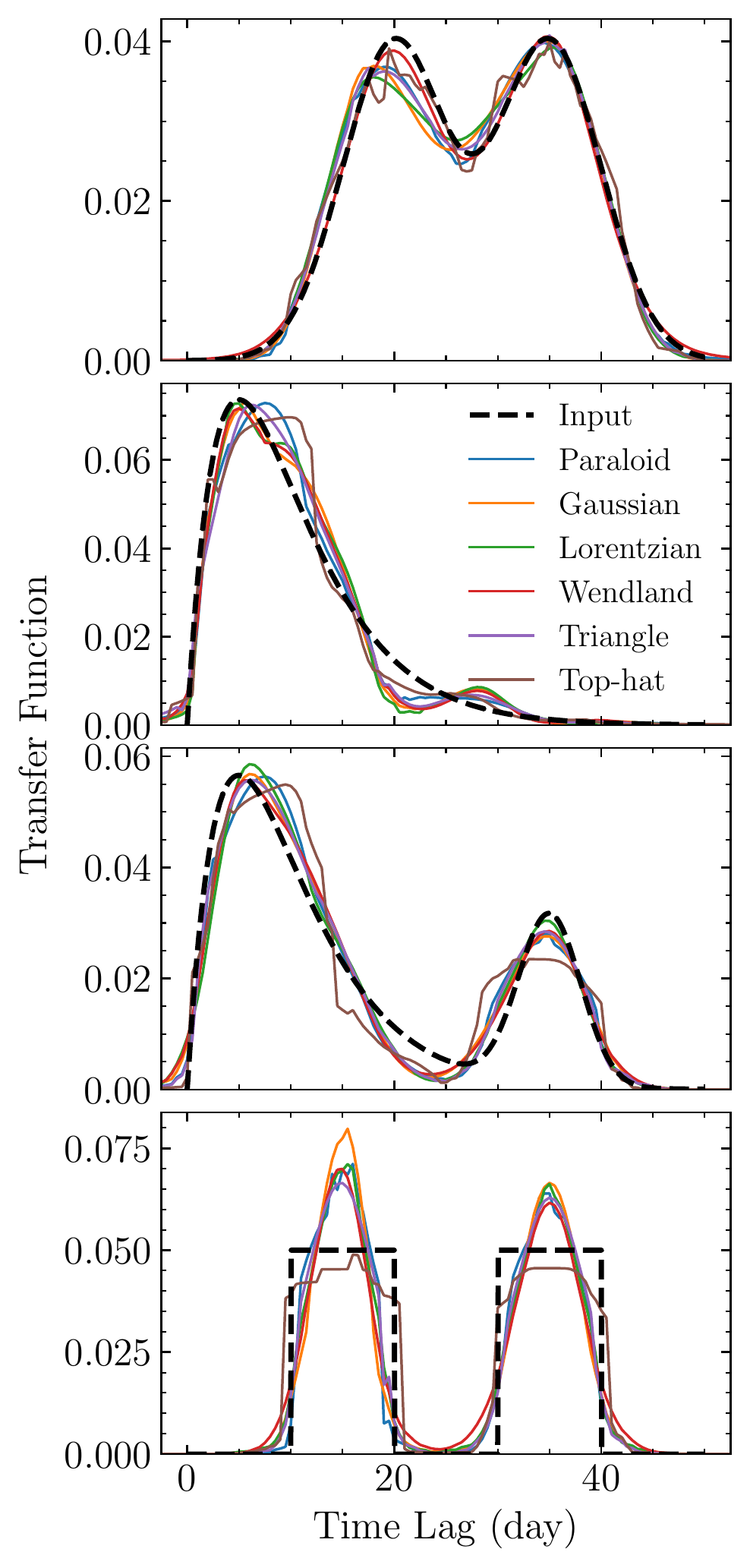}
\caption{\cblue Tests on various shapes of transfer functions with different pixon basis functions.
From top to bottom panels, the shapes of input transfer functions are two highly blending 
Gaussians, a fast-rising peak with a long-descending tail,  a fast-rising peak with a long-descending tail plus a Gaussian, 
and two displaced top hats. The simulated continuum and line light curves are the same as in Figure~\ref{fig_tranfuns_tests}.
The continuum light curves are reconstructed using DRW. }
\label{fig_bases_tests}
\end{figure}

\subsubsection{Continuum Reconstruction Using the DRW Process}
\label{sec_cont_drw}
With the DRW process, the continuum reconstruction is given by (\citealt{Rybicki1992, Li2018})
\begin{eqnarray}
\mathbi{F}_{c}(t) =(\mathbi{Q}^{1/2}\boldsymbol{\xi}_s + \mathbi{\hat s}) + \mathbi{L}(\mathbi{C}_q^{1/2}\boldsymbol{\xi}_q + \mathbi{\hat q}),
\label{eqn_cont_drw}
\end{eqnarray}
where $\mathbi{L}$ is a vector with all unity elements, $\boldsymbol{\xi}_s$ and $\boldsymbol{\xi}_q$ are random variables 
following a normal distribution, and 
\begin{eqnarray}
&&\mathbi{C_q} = (\mathbi{L}^T\mathbi{C}^{-1}\mathbi{L})^{-1},\nonumber\\
&&\mathbi{Q} = [\mathbi{S}^{-1}+\mathbi{N}^{-1}]^{-1},\nonumber \\
&&\mathbi{\hat q} = \mathbi{C_q}\mathbi{L}^T\mathbi{C}^{-1}\mathbi{y}_c,\nonumber\\
&&\mathbi{\hat s} = \mathbi{SC}^{-1}[\mathbi{y}_c - \mathbi{L}(\mathbi{C}_q^{1/2}\boldsymbol{\xi}_q + \mathbi{\hat q})].
\label{eqn_cont_drw_eqs}
\end{eqnarray}
Here $\mathbi{y}_c$ represents the measured continuum light curve, 
$\mathbi{\hat q}$ represents the best estimate for the mean of the continuum light curve, 
$\mathbi{\hat s}$ represents the best estimate for the underlying variation signal,
$\mathbi{N}$ is the covariance matrix of the measurement {\cblue errors}, $\mathbi{S}$ 
is the covariance matrix of the variation signal, and $\mathbi{C}=\mathbi{S}+\mathbi{N}$.
For a DRW process, the covariance depends on time difference $\Delta t$ as 
\begin{equation}
S(\Delta t) = \sigma_{\rm d}^2\exp\left( -\frac{|\Delta t|}{\tau_{\rm d}}\right),
\label{eqn_drw_cov}
\end{equation}
where $\sigma_{\rm d}$ and $\tau_{\rm d}$ are two parameters, 
which represent the standard deviation of {\cblue variations} on a long-term timescale and the typical 
timescale of {\cblue variations}, respectively. 

The parameter set for continuum reconstruction with the DRW process
includes $\sigma_{\rm d}$, $\tau_{\rm d}$, $\boldsymbol{\xi}_s$, and $\boldsymbol{\xi}_q$.
Once $\sigma_{\rm d}$ and $\tau_{\rm d}$ are given, the matrices $\mathbi{S}$, $\mathbi{C}$, $\mathbi{C_q}$, and $\mathbi{Q}$
in Equations~(\ref{eqn_cont_drw}) and (\ref{eqn_cont_drw_eqs}) will be uniquely determined. 
To simplify calculations, we determine the best values of $\sigma_{\rm d}$ and $\tau_{\rm d}$
by only optimizing the posterior probability for the continuum light curve (see \citealt{Li2018}).
In subsequent RM analysis, we then fix $\sigma_{\rm d}$ and $\tau_{\rm d}$ with the best values but 
set $\boldsymbol{\xi}_s$ and $\boldsymbol{\xi}_q$ as free parameters. As such, there is no need 
to calculate the above matrices and their determinants every
step of RM analysis, and therefore the computational speed will be significantly improved. 

Note that in this procedure the continuum reconstruction is already best optimized, and therefore 
the chi square for continuum reconstruction is no longer necessary.

\begin{figure*}
\centering 
\includegraphics[width=0.9\textwidth]{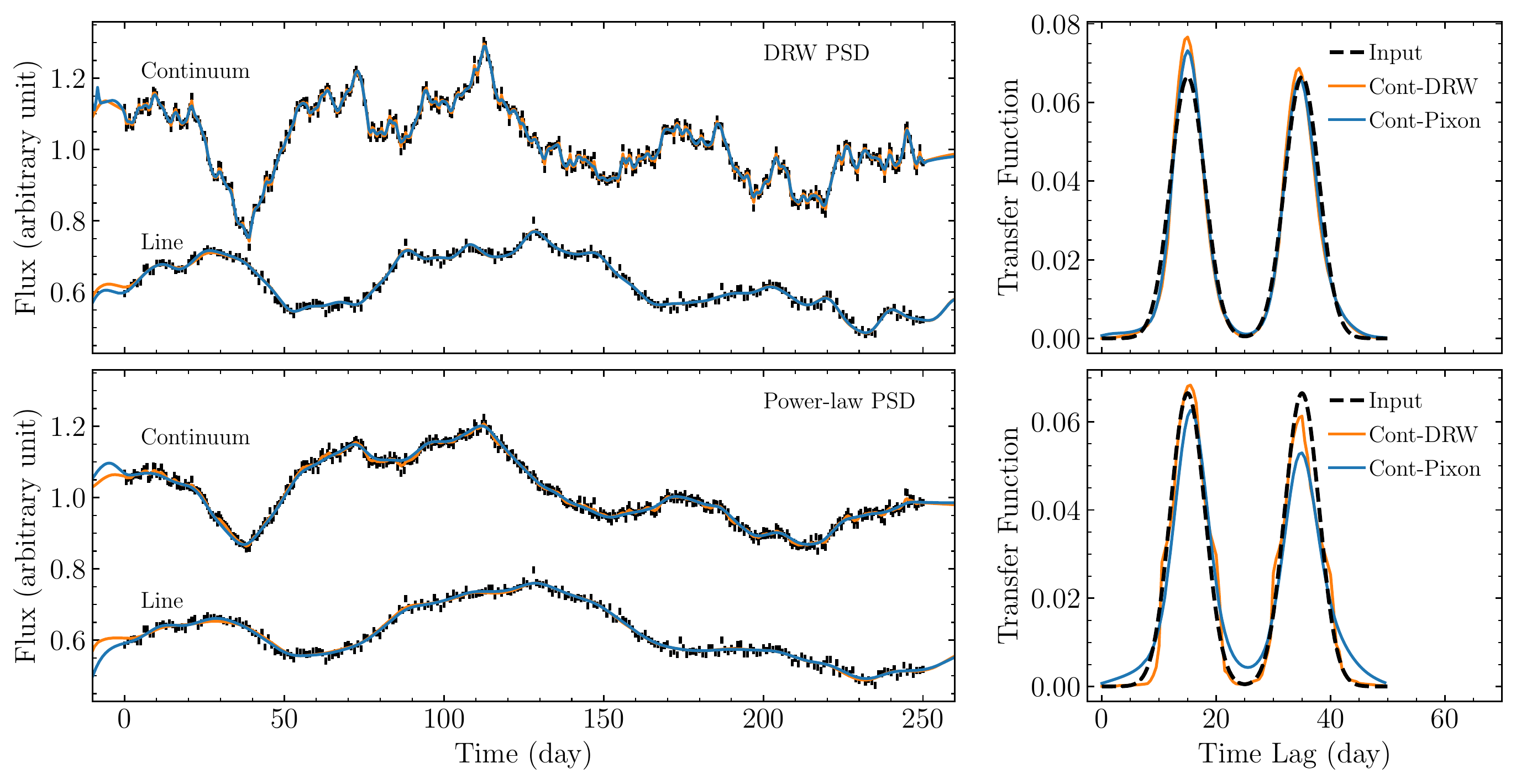}
\caption{Tests with different variation models. Left: simulated light curves using the DRW process in the top panel and 
the power-law PSD in the bottom panel. Right: reconstructions to the input transfer function composed of two displaced Gaussians.
Blue and yellow lines represent the approaches of continuum reconstruction using the pixon concept and DRW process, respectively (see Sections~\ref{sec_cont_pixon} and \ref{sec_cont_drw}). We use the the same random seed so that the global patterns of the the simulated 
light curves are similar. The pixon basis functions are set to Gaussian.
\label{fig_variation_model_tests}}
\end{figure*}

\subsection{Optimization}
\label{sec_opt}
As described above, in the pixon concept, to optimize the posterior probability of Equation~(\ref{eqn_post}), 
we alternatively maximize the exponential term of the posterior probability and meanwhile find the fewest 
allowed number of pixons. As a result, there are two aspects for optimization: determining the best pixon map 
that specifies the pixon size at each pixel and the best solutions for the transfer function. 
This is generally implemented in an iterative manner. First, we presume a pixon map, and with this 
pixon map fixed, we solve for the best solution of the transfer function. Then with the solution 
of the transfer function fixed, we update the pixon map. We iterate this recipe until the termination 
conditions are satisfied.

We rewrite the exponential term of the posterior probability (Equation~\ref{eqn_post}) in logarithm
\begin{equation}
\ln P(I, M|D)\propto -\left(\chi^2 + 2S\right).
\label{eqn_chi_entropy}
\end{equation} 
The expressions for $\chi^2$ and $S$ are slightly different between the two approaches of continuum reconstructions
described in Section~\ref{sec_cont}.
(1) For continuum reconstruction with the pixon concept, the total chi square is the sum of 
\begin{equation}
\chi^2 = \chi^2_l + \chi^2_c.
\end{equation}
Similarly, the total entropy is the sum of 
\begin{equation}
S = S_l + S_c,
\end{equation}
where $S_{l}$ and $S_{c}$ are the entropies for line and continuum reconstruction, respectively.  
(2) For continuum reconstruction with the DRW process, there is no need to 
include the chi square and entropy term for continuum reconstruction, so that $\chi^2 = \chi^2_l$ 
and $S = S_l$. We use a truncated Newton algorithm developed by \cite{Nash1984} to maximize Equation~(\ref{eqn_chi_entropy}).
This needs to calculate gradients of $\chi^2$ and $S$ with respect to parameters.
In the Appendix, we present derivations for these gradients. 

Following \cite{Metcalf1996}, we start with a pixon map that has the largest, uniform pixon size.
This ensures that we compute the large-scale structures in the image before resolving the small-scale structures. 
We then iterate the image and pixon map with the above two-step recipe and gradually reduce the pixon sizes.
To examine whether the pixon size at $i$-pixel is allowed to reduce, we calculate the differential
\begin{equation}
\Delta G = \Delta \delta_i \frac{\partial G}{\partial \delta_i},
\end{equation}
where $G = \chi^2+2S$. In the meantime, the number of degrees of freedom at $i$-pixel also is subject {\cblue to}
a change 
\begin{equation}
\Delta f_i = \Delta \delta_i \frac{\partial f_i}{\partial \delta_i},
\end{equation}
where $f_i$ is defined by Equation~(\ref{eqn_f}). We note that $\Delta G$ can also be deemed to be a change in the number of degrees of 
freedom. In this sense, the condition for {\cblue reducing} the pixon size is (\citealt{Metcalf1996})
\begin{equation}
-\Delta G > \Delta f_i +\sigma \frac{\Delta f_i}{\sqrt{2f_i}},
\end{equation}
where $\sigma$ is a factor to control the sensitivity of reducing pixon sizes, 
$\sqrt{2f_i}$ is the expected standard deviation of the $\chi^2$-distribution with a degree of freedom of $f_i$,
and the minus sign to $\Delta G$ comes from the fact that $G$ decreases while $f_i$ increases with decreasing $\delta_i$.
If the condition  at a pixel is satisfied, the pixon size at that 
pixel is reduced; if not, the previous pixon size is retained.
The pixon map is updated once all pixels are tested, and the overall iteration is terminated if there is 
no further update to the pixon map.

\begin{figure*}[th!]
\centering 
\includegraphics[width=0.9\textwidth]{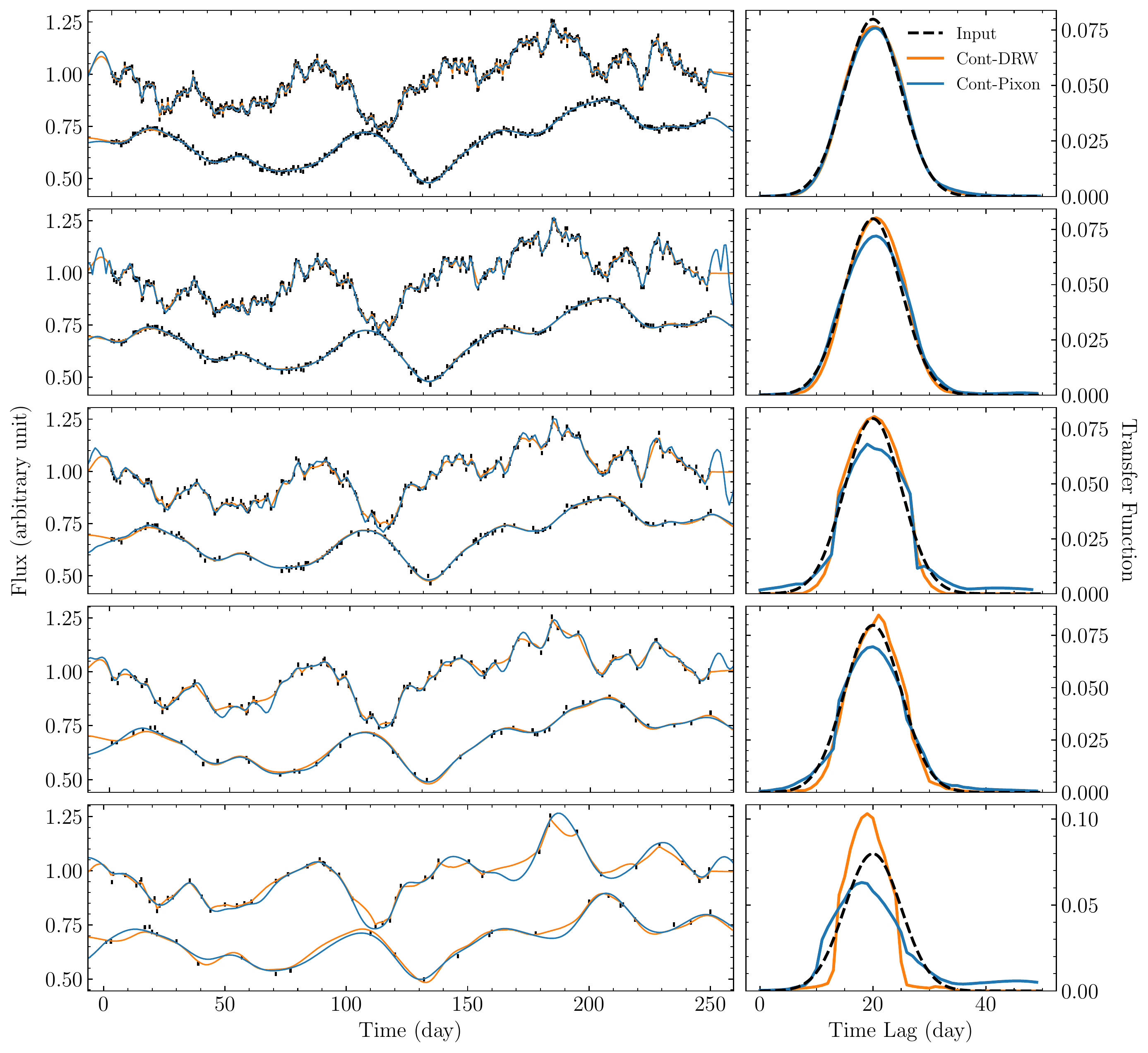}
\caption{Tests with different cadences. Blue and yellow lines represent the approaches of continuum reconstruction
using the pixon concept and DRW process, respectively (see Sections~\ref{sec_cont_pixon} and \ref{sec_cont_drw}).
The pixon basis functions are set to Gaussian.
\label{fig_cadence_tests}}
\end{figure*}

\begin{figure*}[th!]
\centering 
\includegraphics[width=0.9\textwidth]{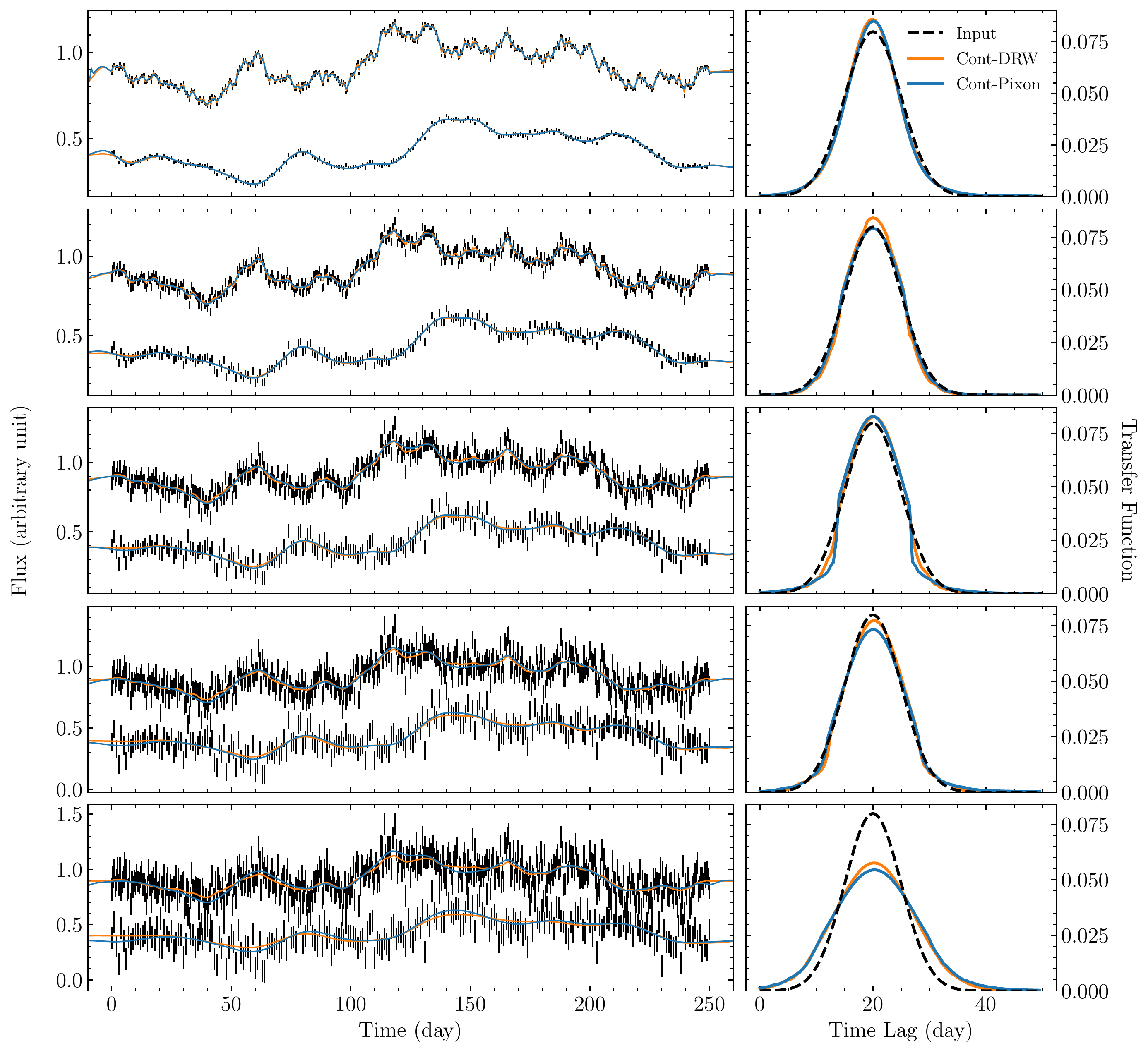}
\caption{Tests with different errors. Blue and yellow lines represent the approaches of continuum reconstruction
using the pixon concept and DRW process, respectively (see Sections~\ref{sec_cont_pixon} and \ref{sec_cont_drw}).
The pixon basis functions are set to Gaussian.
\label{fig_error_tests}}
\end{figure*}

\section{Tests}
We perform basic tests on our pixon-based approach for RM analysis by simulating
a number of artificial light curves with given power spectral density (PSD) models (e.g., \citealt{Li2018mn}).
Unless stated otherwise, the simulations are set up as follows.
We use the DRW {\cblue process (see Equation~\ref{eqn_drw_cov})} to generate light curves, {\cblue which} has a PSD form of%
{\footnote{We note that in the literature PSDs may differ by constant factors, depending on adopted normalizations (\citealt{Deeming1975}).}}
\begin{equation}
P(\nu) = \frac{2\sigma_{\rm d}^2\tau_{\rm d}}{1+(2\pi\tau_{\rm d}\nu)^2}.
\end{equation}
We fix the parameters $\sigma_{\rm d}=0.15$ (arbitrary unit) {\cblue and} $\tau_{\rm d}=50$ days. 
The total time lengths are set to 250 days and the means of generated light curves 
are about one (arbitrary unit; {\cblue subject to} random fluctuations). The {\cblue S/N} ratios are about 100. 
The {\cblue values of the model parameters adopted above} typically result in variation amplitudes of 30\%-50\%. 

\subsection{Validity Tests}
We first test the validity of our pixon-based approach. We randomly generate a continuum light curve with the DRW process. 
The cadence is 0.5 days apart. The light curve of emission line is obtained by convolving the continuum light curve with a Gaussian transfer function.
The Gaussian has a center of 20 days and standard deviation of 5 days. The cadence of the emission-line light curve is 1 day apart.
The left panels of Figure~\ref{fig_validity_tests} plot the generated light curves of the continuum and emission line.
We then use the different pixon basis functions listed in Table~\ref{tab_pixon} to recover the transfer function.
In Figure~\ref{fig_validity_tests}, the top right panel shows results for the case of continuum reconstruction using the pixon concept 
(see Section \ref{sec_cont_pixon}), and the bottom panel is for the case of continuum reconstruction using DRW (see Section \ref{sec_cont_drw}).
The right two panels of Figure~\ref{fig_validity_tests} plot the correspondingly recovered transfer functions.
We can find that all pixon basis functions yield Gaussian-like transfer functions well consistent
with the input.

In Figure~\ref{fig_tranfuns_tests}, we show the recovery to various shapes of input transfer functions, 
including two highly blending Gaussians, a fast-rising peak with a long-descending tail,  a fast-rising 
peak with a long-descending tail plus a Gaussian, and two displaced top hats. We only 
use the DRW process to reconstruct the continuum and the Gaussian pixon basis functions.
As can be seen, the major structures in the input transfer functions are overall well produced, 
although some sharp or discontinuous features are not captured because the pixon concept still favors smoothing solutions.
Nevertheless, the above tests illustrate that the application of the pixon concept to RM analysis is feasible.

We note that the generated light curves in this section are somehow ideal in terms of data cadences. 
Below we will explore the performance of our approach under different simulation configurations.

\begin{figure*}[th!]
\centering 
\includegraphics[width=0.9\textwidth]{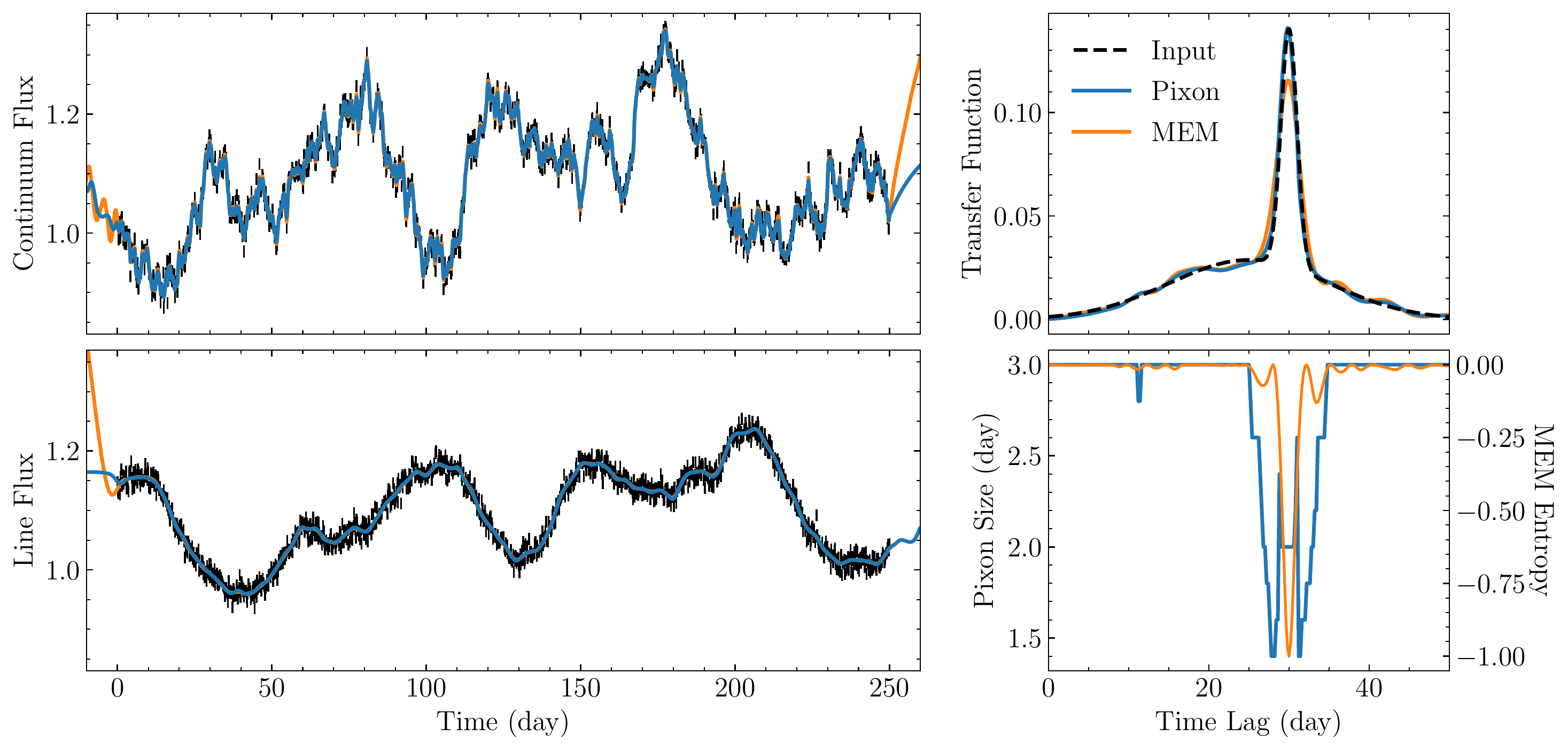}
\caption{A comparison with the MEM on intensively sampled artificial light curves. Left: artificial light curves
of the continuum and emission line. Blue and yellow solid lines represent reconstructions with our pixon method and the MEM, respectively.
Top right: reconstructed transfer functions using our pixon method (blue) and the MEM (yellow).
Bottom right: obtained pixon size distribution with time lag by our pixon method and entropy distribution by the MEM.
The MEM entropy is normalized by the absolute of the minimum value.
\label{fig_comparison}}
\end{figure*}

{\cblue
\subsection{Different Pixon Bases}
Figure~\ref{fig_validity_tests} illustrates the overall consistent recovery to a simple Gaussian transfer function with 
different pixon basis functions. In this section, we test the performance of pixon basic functions
on more complicated transfer functions. In Figure~\ref{fig_bases_tests}, we implement our pixon-based approach 
on the same simulated light curves generated in Figure~\ref{fig_tranfuns_tests} with the pixon basic functions
listed in Table~\ref{tab_pixon}. The continuum light curves are reconstructed using DRW.
The input transfer functions have the same shapes as in Figure~\ref{fig_tranfuns_tests}, namely, 
two highly blending Gaussians, a fast-rising peak with a long-descending tail,  a fast-rising 
peak with a long-descending tail plus a Gaussian, and two displaced top hats.
We can find that all pixon basic functions overall reproduce the major features in the input 
transfer functions.
As mentioned above, the top-hat pixon basic function with hard edges is more inclined to yield 
unsmoothed small-scale features in transfer functions.
We inspect the reconstructed light curves and do not find any significant 
differences with those reconstructed by other pixon basic functions.
Nevertheless, the bottom panel of Figure~\ref{fig_bases_tests} demonstrates that
the top-hat basic function may be superior in capturing very sharp features in transfer functions. 
}

\subsection{Different Variation Models}
To verify whether our approach depends on variation models of light curves, besides the DRW model, we also use a power-law PSD with a form of 
\begin{equation}
P(\nu) = A\left(\frac{\nu}{10^{-2}~{\rm day}^{-1}}\right)^{-\alpha}.
\end{equation}
We set $A=0.3$ (arbitrary unit) and $\alpha=3$. The power-law PSD is forced to flatten to a constant below $5\times10^{-3}~{\rm day}^{-1}$ to 
alleviate the issue of spectral power leakages (\citealt{Uttley2002}). 
The left panels of Figure~\ref{fig_variation_model_tests} plot randomly generated light curves using the DRW process and power-law PSD.
We use the the same random seed in the two PSD models so that the global patterns of the light curves are similar.
As expected, the light curve generated by the DRW process has much more significant short-timescale variations.
The right panels of Figure~\ref{fig_variation_model_tests} show reconstructions to the input transfer function composed of 
two displaced Gaussians using the two approaches for continuum reconstructions (see Sections~\ref{sec_cont_pixon} and \ref{sec_cont_drw}).
The approach based on the DRW process presumes that the continuum variations follow a DRW process, whereas the approach 
based on the pixon concept is less model dependent.
Figure~\ref{fig_variation_model_tests} demonstrates that the two peaks in the input transfer function are overall well reproduced
in both approaches.

\subsection{Different Data Cadences and Errors}
In Figure~\ref{fig_cadence_tests}, we first generate a continuum light curve and then gradually censor some fraction 
of data points randomly to simulate the influence of different cadences. From top to bottom panels of 
Figure~\ref{fig_cadence_tests}, the mean cadences of the continuum light curves are set to
0.5, 0.7, 1.3, 2.3, and 5 days apart and the mean cadences of the emission-line light curves are set to 
1, 1.3, 2.3, 5, and 10 days apart. In the first four cases, the recovered transfer functions are generally 
consistent with the input Gaussian transfer function. In the last case, where the sampling interval (10 days) of the emission-line light 
curve increases to one-half of the mean time lag of the input transfer function (20 days), the obtained 
transfer function displays mild deviations in the peak location, in particular for 
the continuum reconstruction using the pixon concept (see Section~\ref{sec_cont_pixon}). We 
ascribe these deviations to the reason that the information contained in the light curves
does not suffice to constrain the underlying variations appropriately.
Notwithstanding, the peak-like shapes in the transfer function are still retained.

In Figure~\ref{fig_error_tests}, we do testing with different errors. The typical S/Ns 
of the five pairs of generated light curves are set to 100, 40, 20, 13, and 10 from the top to bottom panels of 
Figure~\ref{fig_error_tests}. Again, only the Gaussian pixon basis functions are used.
The obtained transfer functions are in good agreement with the input in both shape and amplitude for the four 
cases with S/N$>$13.  For the case of S/N$=$10, the obtained transfer functions have the same centers of time lags as the input 
but with slightly broader widths.

\section{Comparison with the Maximum Entropy Method}
The pixon method can be regarded as an extension of the MEM in the sense that it allows us to adjust the image prior according to 
the information content of the data. In the MEM, one can also design some forms of default prior images to steer 
the solutions toward those forms of priors as closely as possible under the constraints of the data.
For example, \cite{Horne1994} chose the ``curvature default'' prior, approximately equivalent to 
minimizing curvatures of the solutions (in logarithm). By contrast, the pixon method does not assign default prior images explicitly;
instead, it seeks to smooth the solutions locally as much as the data permit with the fewest number of pixons (namely, minimum complexity). 
It seems that this provides a natural and generic way to implement the principle of {\it Ockham's Razor} (\citealt{Putter1999}),
{\cblue a common rule of thumb in model selections via marginal likelihood maximization.} 
As such, the issue of overresolution or underresolution are minimal in pixon solutions.
Meanwhile, because the pixon method finds the fewest number of pixons, the termination condition in the iterative implementation
can be well specified (see Section~\ref{sec_opt}), which {\cblue involves not only}  the goodness of fit (i.e., $\chi^2$), but also 
the number of pixons. In comparison, the MEM mainly relies on adjusting a weight parameter ($\alpha$), which controls the trade-off between 
the goodness of fit and entropy, to enforce the reduced $\chi^2$ to approximate unity. 
This is one of the major differences between the pixon method and the MEM. 

In Figure~\ref{fig_tranfuns_tests}, we superpose the transfer functions obtained by the MEM for the sake of comparison. 
Here, we use the MEM implementation of \cite{Xiao2018}, which adopts the ``curvature defaults'' following \cite{Horne1994}.
The results obtained by the both approaches are generally consistent with the input transfer functions.
A noticeable difference revealed in the bottom right panel of Figure~\ref{fig_tranfuns_tests} {\cblue and the bottom panel 
of Figure~\ref{fig_bases_tests}} is that 
the pixon method is relatively more capable of producing sharp features.

We further design a test with an extreme transfer function composed of a broad Gaussian and very narrow Gaussian.
The light curves are generated with a high cadence (0.2 days apart) to provide adequate information to 
recover the narrow Gaussian component. We run the pixon method and the MEM on the light curves and show the results 
in Figure~\ref{fig_comparison}. The broad Gaussian component in the transfer function 
is well reproduced by both the pixon method and the MEM, although there appear to be some very minor ripple-like
features. However, for the narrow Gaussian component, 
the MEM seems to yield a slightly larger width, whereas the pixon method obtains a remarkably consistent width.
In the bottom right panel of Figure~\ref{fig_comparison}, we show the pixon size distribution with time lag obtained by 
the pixon method and the entropy distribution obtained by the MEM. As \cite{Horne1994} pointed out, 
the entropy with ``curvature defaults'' results in preference to solutions with Gaussian peaks ($\propto \exp(-\tau^2/2)$) and 
exponential tails ($\propto \exp(-\tau)$). The former have constant curvatures, and the latter have zero curvatures, 
both of which maximize the entropy.  This might be the reason responsible for the resulting slightly larger width of 
the narrow Gaussian component. For the pixon results, we can find that 
the pixon size reaches a minimum around time lags where the transfer function changes the most rapidly, 
instead of around the peak of the transfer function. This is consistent with the expectation that 
high resolutions are required to recover rapid changing features in the transfer function.

\section{Discussion and Conclusion}
We adapt the pixon algorithm initially proposed for image reconstruction by \cite{Pina1993} to 
RM analysis and develop a generic framework to implement the algorithm.
The pixon method uses pixons (instead of pixels) as the basic unit, which are able 
to adjust {\cblue pixon} sizes to achieve locally optimal resolutions according to 
the information content provided by the data. Within a pixon, the pixon algorithm smooths the solutions 
as much as the data allow. The terminated criterion of the pixon method is to find the 
fewest number of pixons that still adequately fit the data. This naturally 
obeys the principle of {\it Ockham's Razor} (\citealt{Putter1999}).
As such, the pixon method optimizes solutions not only by testing the goodness of fit but also by 
reducing the complexity to be optimal. The issue of overresolution or underresolution 
of the solutions is also significantly alleviated.
{\cblue In this sense}, the pixon method is flexible to optimize the complexity of the solutions
{\cblue and can be regarded as a subcase of marginal likelihood maximization.}

We perform a number of simulations to illustrate the validity of our pixon-based approach for 
RM analysis. We also compare the pixon method with the widely used MEM on simulated light curves 
(see Figures~\ref{fig_tranfuns_tests}
and \ref{fig_comparison}) and find that both the approaches generally give consistent results 
to the input transfer functions. However, in some cases, the pixon method performs better in 
producing sharp features in transfer functions.

There are several potential improvements to be made to our pixon-based RM analysis in future:
(1) Currently, we only apply the pixon algorithm to velocity-unresolved RM analysis. 
It is straightforward to extend the current framework to velocity-resolved cases. 
Accordingly, the pixon basis should be functions of time lag and velocity. 
{\cblue This} might need some elaboration on the extensions of pixon basis functions along time lag and velocity directions.
(2) We adopt the truncated Newton algorithm developed by \cite{Nash1984} to seek the 
optimized solutions. This algorithm does not require us to supply the Hessian information. 
In our pixon method, it is easy to analytically calculate the Hessian matrix; therefore, 
other optimization algorithms that make use of the Hessian information are worthwhile to 
apply to expedite the computational efficiency.
(3) Our present pixon method is unable to estimate the uncertainties of the obtained solutions.
A Markov Chain Monte Carlo may help to achieve this purpose.

We finally mention that the present approach applies not only to RM analysis of emission lines but also 
to RM analysis on  multiband continuum light curves. {\cblue We developed a software package \texttt{PIXON}
for our pixon method and made it publicly available at \url{https://github.com/LiyrAstroph/PIXON}.}

\acknowledgements
{
This research is supported in part by the National Key R\&D Program of China (2016YFA0400701); 
by grant Nos. NSFC-11833008, -11991051, and -11991054 from the National Natural Science Foundation of China;
and by the China Manned Space Project with No. CMS-CSST-2021-A06 and CMS-CSST-2021-B11.
Y.-R.L. acknowledges financial support from the National Natural Science Foundation of China through grant No. 11922304, 
from the Strategic Priority Research Program of CAS through grant No. XDB23000000, and from the Youth Innovation 
Promotion Association CAS. X.M. acknowledges financial support from the National Natural Science Foundation of China through grant No. 12003036.
}

\appendix 
In this {\cblue appendix}, we present calculations of the gradients required to supply to the truncated Newton algorithm.
The parameter set to be solved includes $I^{(p)}$, $F_c^{(p)}$, $\boldsymbol{\xi}_s$, and $\boldsymbol{\xi}_q$. 
We derive the differentials of $\chi_l^2$ and $\chi_c^2$ first with respect to the parameter set and then with 
respect to the pixon size. To ensure that the reconstructed image is non-negative, we use logarithm of the pseudo-image $I^{(p)}$.

The derivative of $\chi_l^2$ with respect to $\ln I_i^{(p)}$ is 
\begin{equation}
\frac{\partial \chi_l^2}{\partial \ln I_i^{(p)}}
=  2I_i^{(p)}\sum_j\frac{R_{l, j}}{\sigma_{l, j}^2} \frac{\partial F_{l, j}}{\partial I_i^{(p)}},
\end{equation}
where the residual $R_{l, j}$ is given by
\begin{equation}
R_{l, j} = F_{l, j} - D_{l, j}.
\end{equation}
The derivative of $F_{l, j}$ with respect to $I_i^{(p)}$ is written as
\begin{equation}
\frac{\partial F_{l, j}}{\partial I_i^{(p)}} = \int \frac{\partial I(\tau)}{\partial I_i^{(p)}} F_{c}(t_j-\tau) d\tau.
\end{equation}
We note that 
\begin{equation}
 \frac{\partial I(\tau)}{\partial I_i^{(p)}} = K_\tau\left(\frac{\tau - \tau_i}{\delta_\tau}\right),
\end{equation}
which represents the magnitude of the pixon function of pixel $\tau$ at pixel $i$.
As a result, we have 
\begin{equation}
\frac{\partial F_{l, j}}{\partial I_i^{(p)}} = \int K_\tau\left(\frac{\tau - \tau_i}{\delta_\tau}\right) F_{c}(t_j-\tau) d\tau.
\end{equation}
The derivative of the entropy $S_l$ is 
\begin{eqnarray}
\frac{\partial S_l}{\partial \ln I_i^{(p)}}=-\alpha\frac{\partial}{\partial \ln I_i^{(p)}}\sum_j \frac{I_j}{I_{\rm tot}}\ln\frac{I_j}{I_{\rm tot}}
=-\alpha\frac{I_i^{(p)}}{I_{\rm tot}}\sum_j K_{ji}\left(1+\ln\frac{I_j}{I_{\rm tot}}\right),
\end{eqnarray}
where 
\begin{equation}
 K_{ji} = \frac{\partial I_j}{\partial I_i^{(p)}} = K_j\left(\frac{\tau_j - \tau_i}{\delta_j}\right).
\end{equation}
The derivative of $\chi_l^2$ with respect to the continuum pseudo-image $F_{c, i}^{(p)}$ is
\begin{eqnarray}
\frac{\partial \chi_l^2}{\partial F_{c, i}^{(p)}} &=& 2 \sum_j \frac{R_{l, j}}{\sigma_{l, j}^2} \frac{\partial F_{l, j}}{\partial F_{c, i}^{(p)}} \nonumber\\
& = & 2 \sum_j \frac{R_{l, j}}{\sigma_{l, j}^2} \int I(\tau) \frac{\partial F_c(t_j-\tau)}{\partial F_{c, i}^{(p)}}d\tau \nonumber\\
& = &  2 \sum_j \frac{R_{l, j}}{\sigma_{l, j}^2} \int I(\tau) K_{m}\left(\frac{t_j - t_i - \tau}{\delta_{m}}\right) d\tau,
\end{eqnarray}
where $K_{m}$ and $\delta_{m}$ are the pixon function and size appropriate at $t_j-\tau$, respectively.
Since we use a uniform pixon size for continuum reconstruction, $K_{m}$ and $\delta_{m}$ are the same over all pixels.
Using the chain rule for differentiation, the derivatives of $\chi_l^2$ with respect to $\boldsymbol{\xi}_q$ and $\boldsymbol{\xi}_q$ are given by 
\begin{eqnarray}
\frac{\partial \chi_l^2}{\partial \boldsymbol{\xi}_s} &=& 
\frac{\partial \mathbi{F_c}}{\partial \boldsymbol{\xi}_s} \frac{\partial \chi_l^2}{\partial \mathbi{F_c}}\nonumber\\
&=& 2\mathbi{Q}^{1/2} \sum_j \frac{R_{l, j}}{\sigma_{l, j}^2} \frac{\partial F_{l,j}}{\partial \mathbi{F_c}}\nonumber\\
&=& 2\mathbi{Q}^{1/2} \sum_j \frac{R_{l, j}}{\sigma_{l, j}^2} \int I(\tau)\frac{\partial F_c(t_j-\tau)}{\partial \mathbi{F_c}} d\tau\nonumber\\
&=& 2\mathbi{Q}^{1/2} \sum_j \frac{R_{l, j}}{\sigma_{l, j}^2} I(t_j-\mathbi{t}).
\end{eqnarray}
and 
\begin{eqnarray}
\frac{\partial \chi_l^2}{\partial \boldsymbol{\xi}_q} &=&
\frac{\partial \mathbi{F_c}}{\partial \boldsymbol{\xi}_q}  \frac{\partial \chi_l^2}{\partial \mathbi{F_c}}
=2\left[(\mathbi{L} - \mathbi{SC}^{-1}\mathbi{L})\mathbi{C}_q^{1/2}\right]^T\sum_j \frac{R_{l, j}}{\sigma_{l, j}^2} I(t_j-\mathbi{t}),
\end{eqnarray}
where {\cblue the} superscript ``$T$'' denotes transposition and $\mathbi{t}$ is a vector of time on which the continuum is reconstructed.
$\chi_c^2$ does not depend on $I^{(p)}$ and thus $\partial \chi_c^2/\partial I_i^{(p)}=0$. 
The derivative of $\chi_c^2$ with respect to the continuum pseudo-image $F_{c, i}^{(p)}$ reads
\begin{eqnarray}
\frac{\partial \chi^2_c}{\partial F_{c, i}^{(p)}} = 2 \sum_j \frac{R_{c, j}}{\sigma_{c, j}^2} \frac{\partial F_{c, j}}{\partial F_{c, i}^{(p)}}
= 2 \sum_j \frac{R_{l, j}}{\sigma_{l, j}^2} K_j\left(\frac{t_j - t_i}{\delta_j}\right),
\end{eqnarray}
where the residual $R_{c, j}$ is given by
\begin{equation}
R_{c, j} = F_{c, j} - D_{c, j}.
\end{equation}

For the derivative of $\chi_l^2$ with respect to the pixon size, we have 
\begin{eqnarray}
\frac{\partial \chi_l^2}{\partial \delta_i} &=&  \sum_j \frac{2R_{l, j}}{\sigma_{l, j}^2}\int \frac{\partial I(\tau)}{\partial \delta_i} F_c(t_j-\tau) d\tau
= \frac{\partial I(\tau_i)}{\partial \delta_i}\sum_j \frac{2R_{l, j}}{\sigma_{l, j}^2} F_c(t_j-\tau_i),
\end{eqnarray}
where 
\begin{equation}
\frac{\partial I(\tau_i)}{\partial \delta_i} = \int \frac{\partial K_i}{\partial \delta_i} I^{(p)}(y) dy.
\end{equation}
The derivative of the entropy $S_l$ with respect to the pixon size is
\begin{eqnarray}
\frac{\partial S_l}{\partial \delta_i}&=&-\alpha\frac{\partial}{\partial \delta_i}\sum_j \frac{I_j}{I_{\rm tot}}\ln\frac{I_j}{I_{\rm tot}} 
= -\frac{\alpha}{I_{\rm tot}}\sum_j \left(1 + \ln\frac{I_j}{I_{\rm tot}} \right) \frac{\partial I_j}{\partial \delta_i}
= -\frac{\alpha}{I_{\rm tot}}  \left(1 + \ln\frac{I_i}{I_{\rm tot}} \right) \frac{\partial I_i}{\partial \delta_i},
\end{eqnarray}
where we neglect the change of $\alpha$ with the pixon size.
The derivative of the entropy $S_c$ with respect to the pixon size can be written similarly.

\end{document}